# Non-local Chemical Potential Modulation in Topological Insulators Via Electric Field Driven Trapped Charge Migration

Yasen Hou[1], Rui Xiao[1], Senlei Li[1], Lang Wang[1], Dong Yu[1*]

*[1]Department of Physics, University of California, Davis, California 95616, USA*

*\*e-mail: yu@physics.ucdavis.edu*

**Topological insulators (TIs) host unusual surface states with Dirac dispersion and helical spin texture and hold high potentials for novel applications in spintronics and quantum computing. Control of the chemical potential in these materials is challenging but crucial to realizing the hotly pursued exotic physics, including efficient spin generation[1,2], Majorana Fermions[3-5], and exciton condensation[6,7]. Here we report a simple and effective method that can *in-situ* tune the chemical potential of single-crystal $Bi_{2-x}Sb_xSe_3$ nanoribbons, with a magnitude significantly larger than traditional electrostatic gating. An electric field parallel to a device channel can shift the chemical potential across the Dirac point, both inside and outside the channel. We attribute this non-local reversible modulation of chemical potential to electric-field-induced charge hopping among defect states, further supported by photocurrent mapping. Our approach enables engineering chemical potential distributions in TIs and opens up tremendous opportunities for investigating fundamental transport mechanisms of charge and composite particles in these materials.**

Manipulating chemical potentials in materials plays a crucial role in their optoelectronic applications and is specifically important in TIs because unintentional doping in these materials creates a key obstacle to displaying their desired unique spin and charge transport properties. Chemical doping[8,9], molecule adsorption[9,10], electron or alpha particle irradiation[11,12], potassium deposition[13], and controlled structural deformation[14] have been employed to bring the chemical potential into the TI's bulk bandgap. Nevertheless, for both fundamental studies and electronic applications, a dynamical method of adjusting the chemical potential during device operation is desired, but by far is largely limited to electrostatic gating[15,16]. Back gating is known to be efficient only at influencing the carrier concentration at the TI bottom surface[17], while the fabrication of top gates involves sophisticated processes and tends to degrade the material properties[18,19]. Recently photo-gating has also been demonstrated in $(Bi,Sb)_2Te_3$ devices, but it requires interfacing an optically thin TI layer with a particular substrate such as $SrTiO_3$[20,21].

Charge traps can significantly affect electronic properties, as manifested in nonvolatile memory[22-24] and persistent photoconductivity[25-29]. TI surfaces are vulnerable to various kinds of defects, especially when placed in ambient environments[18,30]. While these defect states are previously largely viewed as detrimental to charge transport, here we show that the active manipulation of the trapped charges offers a powerful way to control the chemical potential. Compared to previous methods, our approach neither introduces external defects as in chemical doping, nor requires creating a foreign interface as in electrostatic or photo-gating. Cooling the devices under electric field results in intrinsic TIs, which demonstrate ambipolar conduction and long photocurrent decay length indicating exciton condensation.

Single-crystal $Bi_{2-x}Sb_xSe_3$ nanoribbons were grown by chemical vapor deposition (CVD)[6,31] with x ranging from 0.15 to 0.4. Field effect transistors (FETs) incorporating single nanoribbons were then fabricated with standard e-beam lithography. Multiple Ohmic Ti/Au top contacts were made as shown in Fig. 1b. After a longitudinal electric field was maintained in the device for a few minutes, the channel conductance gradually changed by a few percent (Extended Data Fig. 1). More surprisingly, the electric field not only changed the conductance in the channel where it was applied, but also outside. In device #1, the conductance in channel CD ($G_{CD}$) increased by 7.5% in hundreds of seconds at 300 K, when a bias was applied in channel AB ($V_{AB}$) 232 μm away (Fig. 1c). After $V_{AB}$ was removed, $G_{CD}$ decreased to the initial value in a similar time. The reverse of the electric field direction led to the opposite change in conductance. Though the conductance change was highly reproducible in all tested $Bi_{2-x}Sb_xSe_3$ nanoribbon devices, pure $Bi_2Se_3$ nanoribbon devices did not show such a behavior likely caused by their high conductivity.

The change rate of the non-local conductance sensitively depended on temperature. At 175 K, the response took more than $10^4$ s (Fig. 1d). After cooling the device under a fixed bias from room temperature to 77 K, the conductance change was maintained indefinitely even after the applied electric field was removed. Furthermore, at low temperature the conductance could no longer be modulated by a longitudinal electric field. Nevertheless, conductance was recovered to the original level if the device was warmed to room temperature. Several cooling/warming cycles were then performed on device #2 with different biasing conditions during cooling (Fig. 2). Gate responses confirmed that the conductivity change originated from the chemical potential (or Fermi level $E_F$) shift. The *n*-type device exhibited strong gate dependence (Fig. 2e, h) when cooling with all contacts floating. Channel AB appeared to be more *n*-type than CD, likely because of slight spatial doping variation during the sample growth. $G_{CD}$ reached a minimum at $V_g = -73$ V, indicating that $E_F$ was tuned across the Dirac point by $V_g$. However, after cooling with $V_{DB} = 5$ V, $G_{AB}$ increased more than 10 times and completely lost gate response, while the minimum point in $G_{CD}$ in the gate scan shifted close to $V_g = 0$ V (Fig. 2d, g). Cooling with $V_{AB} = 1$ V shifted $E_F$ in opposite directions (Fig. 2f, i) and made channel AB ambipolar. The electric field can thus either raise or lower $E_F$ depending on the field direction and the location. $E_F$ rises beyond the low potential contact, and falls in the channel and beyond the high potential contact. The field-controlled bi-directional $E_F$ shift enables design of biasing conditions to engineer spatial distributions of $E_F$, for instance, to create $n^+$-$n$ and $p$-$n$ junctions in TIs. Furthermore, the chemical potential manipulation achieved with the longitudinal field is significantly more effective than conventional electrostatic gating. The conductivity can be tuned by more than 10 times via field cooling (Fig. 2d), while sweeping $V_g$ over a range of 250 V across the 300 nm oxide only leads to a maximum 25% conductivity change.

To gain further insights in the mechanism of the non-local conductance change, we employ scanning photocurrent microscopy (SPCM), where photocurrent is measured as a focused laser is raster scanned over the device plane to form a 2D map. Photocurrent is generated if the locally injected electrons and holes are separated in the presence of an electric field. As a result, this method provides information about the local electric field distribution originated from doping gradient[32]. The experiment was carried out at 200 K in order to slow down the charge migration to allow systematic examination by photocurrent imaging. Device

#1 was first cooled with all contacts floating to 200 K. A photocurrent image showed random photocurrent spots with small magnitude (Fig. 3a), which had been reported before and were attributed to the random fluctuation of chemical potential in TIs[33,34]. Then $V_{CB} = 4$ V was held for 3 hrs to create large doping gradient. After $V_{CB}$ was reduced to zero at t = 0 s, SPCM images were taken consecutively to examine the evolution of the electric field distribution. At t = 10 s, a strong negative photocurrent peak appeared close to contact B (Fig. 3a). Over time, this peak moved towards contact C and gradually became broader and weaker. After 12000 s, the photocurrent image was recovered to that before the bias was applied. $V_{CB} = -4$ V yielded similar results, with both the polarity and the shift direction of the photocurrent peak reversed (Fig. 3b). The peak position follows well with the square root of time (Fig. 3c), as expected from a diffusion process[32]. The fitting yields a diffusion constant of $2.3 \times 10^{-7}$ cm$^2$/s at 200 K. Similar SPCM measurements performed at 300 K showed the same trend with the recovery process taking less than 180 s (Extended Data Fig. 2).

We now discuss the mechanism of the above phenomena. Non-local conductance change (Fig. 1), $E_F$ tuning (Fig. 2), and photocurrent peak shift (Fig. 3) can all be understood by the electric field induced migration of localized charges, such as trapped electrons or ions. These localized charges do not directly contribute to electrical conduction but provide local electrostatic gating and enable $E_F$ shift. We now exclude a couple of possible origins of the localized charges. First, Au migration from the contacts onto the TI surface[35,36] does not explain the observations, since TI devices fabricated with Au-free contacts exhibited similar non-local $E_F$ shift effects (Extended Data Fig. 3). Second, Se vacancies or excess Se atoms at the TI surface unlikely account for the observed fast diffusion, as atoms or ions usually migrate much slower[37-39] in solids. In addition, in our experiment the non-local $E_F$ tuning occurred beyond the contacts, while surface ion migration was expected to be blocked at the metal interface as observed previously[32]. Spatially resolved energy dispersive X-ray spectroscopy (EDS) did not show any evidence of ion migration after biasing (Extended Data Fig. 4).

The most reasonable mechanism is field-driven migration of electrons (or holes) trapped at defect states. Photocurrent mapping shows that trapped charges build up at the low potential contact (Fig. 3a, b), clearly indicating the trapped charges are positively charged, i.e., holes. Our experimental observations can then be understood by drift of trapped holes under electric field to contact and further diffusion beyond this contact (Fig. 1a). In contrast to ions, the hole hopping among traps is not significantly affected by metal contacts, in agreement with our experimental observation. Though the defect states are localized, charge can hop among trap states under an external electric field. We note that this picture is in agreement with a recent theoretical prediction that trapped electrons in the TI surface defects states can be highly mobile[40]. Our observation that photoexcitation at 77 K led to persistent conductance increase (Extended Data Fig. 5) provides further evidence: though the migration of trapped holes is suppressed at this temperature, photoexcited hot charge carriers can be captured by the traps and lead to local $E_F$ modulation[25-29]. This photo-assisted trapping can be used to design and control chemical potential distributions via optical writing at low temperature. Our preliminary test indeed confirmed that scanning a high-power laser over an area of the sample raised $E_F$ and created $n$-$n^+$ junctions at the edges of the writing area, as visualized by SPCM (Extended Data Fig. 6).

To more quantitatively understand this process, we consider a 1D model describing diffusion-drift along the nanoribbon axis, where the trapped hole density $n(x,t)$ follows the continuity equation $\frac{dn}{dt} = \mu E \frac{dn}{dx} - D \frac{d^2n}{dx^2}$. Here $E$ is the electric field along the device, $\mu$ is the mobility, and $D$ is the diffusion constant. The simulated $n$ at the center of probing channel CD follows well the measured conductance change (Fig. 1d) at all temperatures (see more details in Methods). The simulation also agrees with the bias dependence as in Fig. 1e (Extended Data Fig. 7). $D$ extracted from the model is $2.1 \times 10^{-6}$ cm$^2$/s at 300 K, corresponding to a hole hopping mobility of $8.0 \times 10^{-5}$ cm$^2$/Vs. $D$ is temperature sensitive and follows well an Arrhenius behavior with an activation energy of 0.179 eV (Fig. 1e). The $D$ value extracted at 200 K is $2.6 \times 10^{-7}$ cm$^2$/s and agrees well with that extracted from the photocurrent peak shift (Fig. 3c).

The $E_F$ tunability achieved by the longitudinal electric field is stronger in comparison with conventional back-gating as shown in Fig. 2, presumably because of the proximity of the localized charge to the conduction channel and its effectiveness at both top and bottom TI surfaces. Consequently, field cooling provides a practical and efficient way to suppress bulk conduction and reveal the non-trivial surface charge transport of the TI materials. For example, we have recently demonstrated millimeter-long transport of photoexcited carriers up to 40 K in $Bi_{2-x}Sb_xSe_3$ nanoribbons[6], which provided key experimental evidence for theoretically predicted topological exciton condensate[7,41]. Such efficient transport of photocarriers is only observed when $E_F$ is close to the Dirac point. Here by cooling with different biasing conditions, we can selectively control the photocarrier transport distance in different channels and turn on or off the exotic effect at will. To demonstrate, we started with a device (device #1) with $E_F$ too high above the Dirac point to exhibit the non-local photocurrent. When cooled to 77 K with all contacts floating, the device only showed localized and weak photocurrent near the contacts, originating from photo-thermoelectric effects (Fig. 4a). However, after cooling with $V_{BC} = 4$ V, photocurrent was much stronger and extended far beyond the contacts, in the region to the right of contact C (Fig. 4b). This is consistent with $E_F$ lowering in this region, as the electric field sweeps trapped holes out of the channel to the left of contact C. Photocurrent images taken with different connection configurations are all in perfect agreement with the expected $E_F$ distribution as shown in the bottom diagram of Fig. 4b. Similarly, when cooled with $V_{CD} = 0.1$ V, long photocurrent decay was observed to the left of contact C as expected (Fig. 4c). $V_g$ alone cannot turn on the photocurrent without the use of longitudinal electric field.

The electric field can also be used to turn off the photocurrent in the desired channel of the device. This time we started with a different device (device #3), in which $E_F$ was initially close to the Dirac point as confirmed by gate scans (Extended Data Fig. 8). Extended photocurrent signal was observed on the entire device when cooling with all contacts floating (Fig. 4d). Fitting of the photocurrent cross sections shows a photocurrent decay length ($L_d$) up to 260 μm at 77 K (Extended Data Fig. 9). Cooling with $V_{CE} = -4$ V increased $E_F$ to the right of contact C and led to vanishing photocurrent in this region (Fig. 4e). Cooling with $V_{CE} = 4$V led to vanishing photocurrent to the left of contact C (Fig. 4f). The field effects therefore can be used to engineer the $E_F$ distribution in a TI device, to potentially control the critical temperature of exciton condensation and the coherence length of excitons in selected regions.

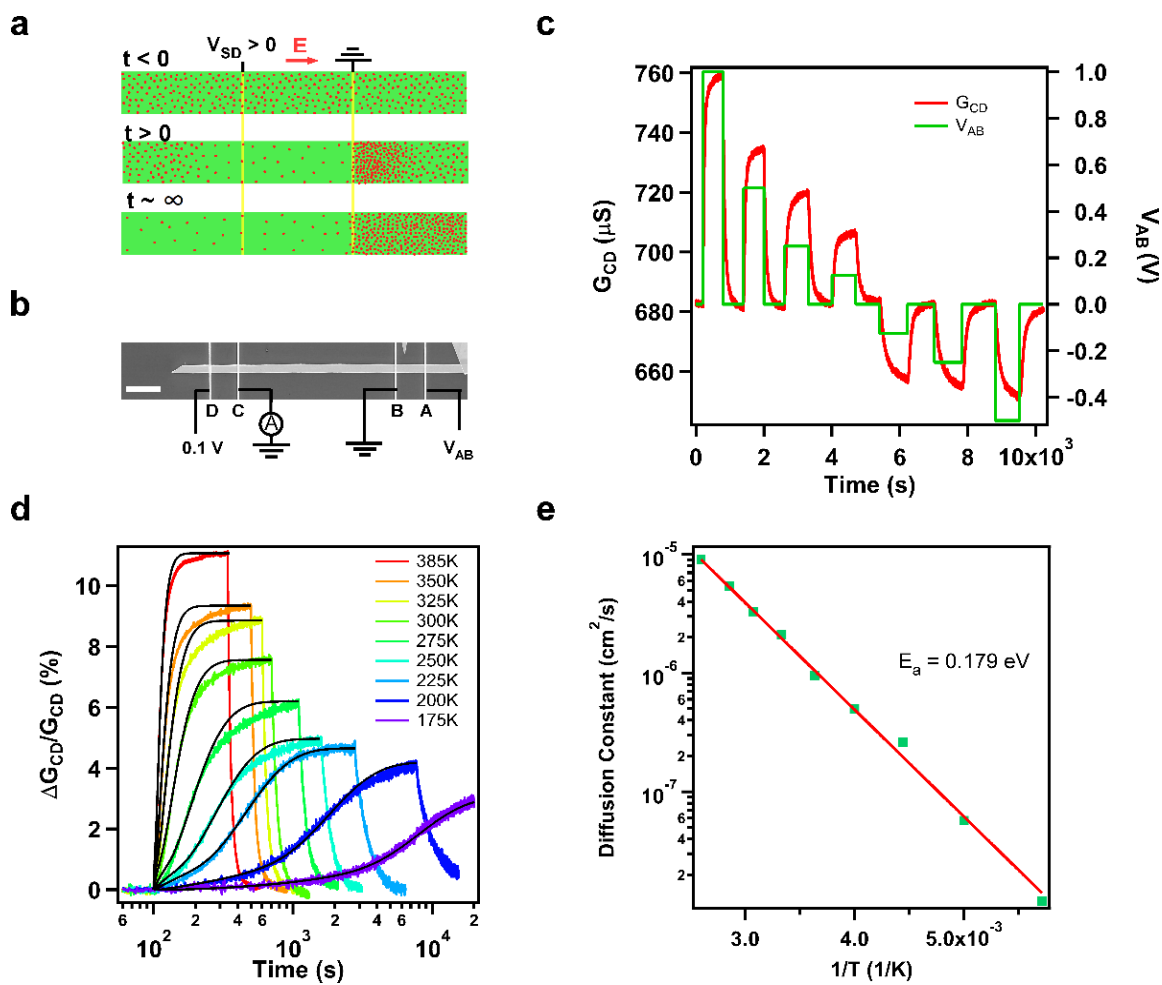

**Fig. 1: Temperature dependent electric-field-driven non-local conductance tuning. a**, Schematic of trapped charge migration with electric field applied after $t = 0$ in the middle section of the device. Red dots represent charges trapped at the defect states. **b,** SEM image of a TI nanoribbon device (device #1) with four top metal contacts. Scale bar denotes 50 μm. **c,** $G_{CD}$ and $V_{AB}$ as a function of time at 300 K. $G_{CD}$ was monitored by a small bias on channel CD. **d,** Time trace of percentage change in $G_{CD}$ in response to $V_{AB}$ at various temperatures. $V_{AB} = 0.5$ V was turned on at 100 s for all temperatures and off when $G_{CD}$ reached saturation. Black curves are simulated responses using a diffusion model. **e,** Semi-log plot of diffusion constant extracted from **d** as a function of inverse temperature. The red curve is fitting by an Arrhenius equation with an activation energy $E_a = 0.179$ eV.

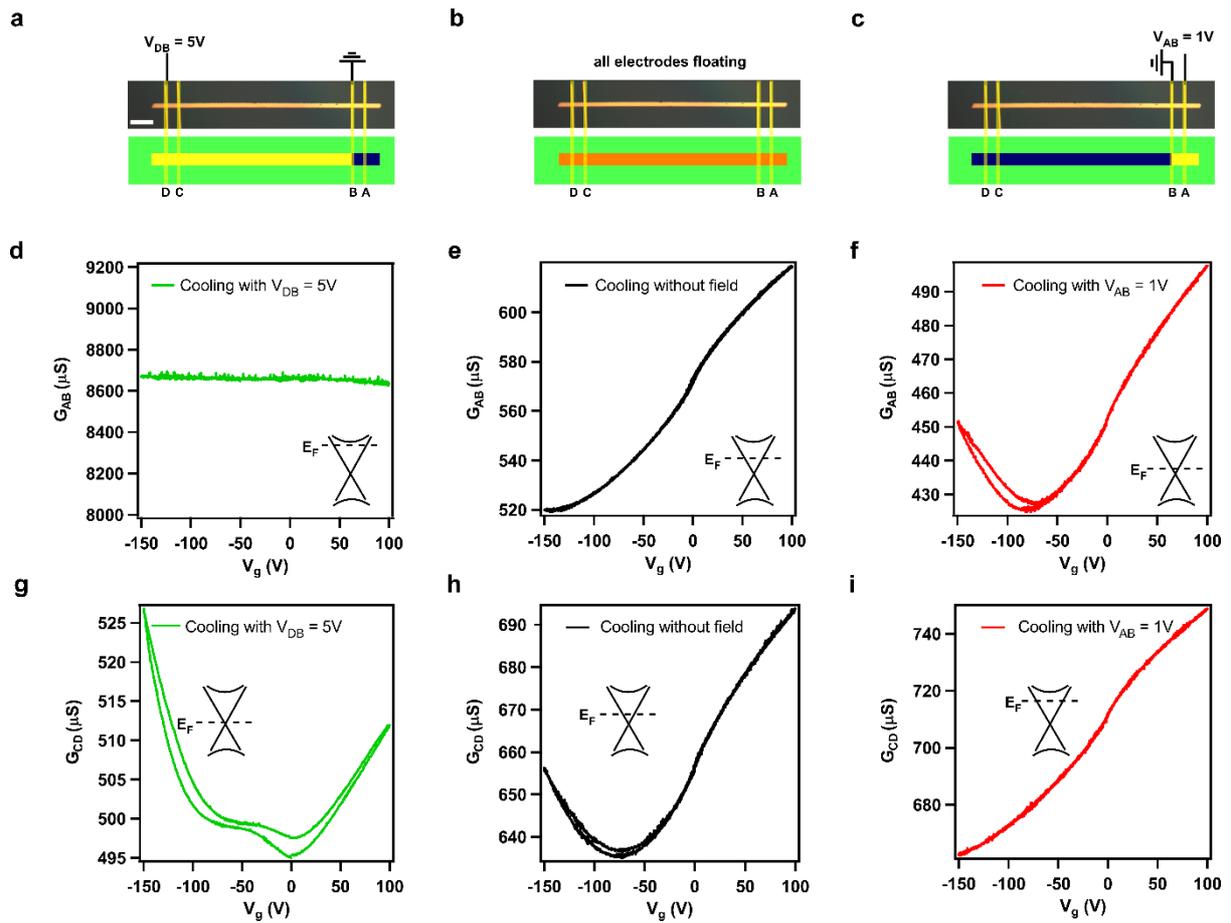

**Fig. 2: Field cooling effects on conductance and gate response. a-c,** Optical image of device #2 and schematics showing three biasing conditions. Scale bar denotes 50 μm. The color codes indicate the conductivity change after cooling to 77 K: orange, no change; yellow, less *n*-type; dark blue, more *n*-type. **d-i,** Gate responses of channels AB (**d-f**) and CD (**g-i**) at 77 K, respectively, after cooling under different biasing conditions. Insets: energy diagrams showing $E_F$ modulation in the channel.

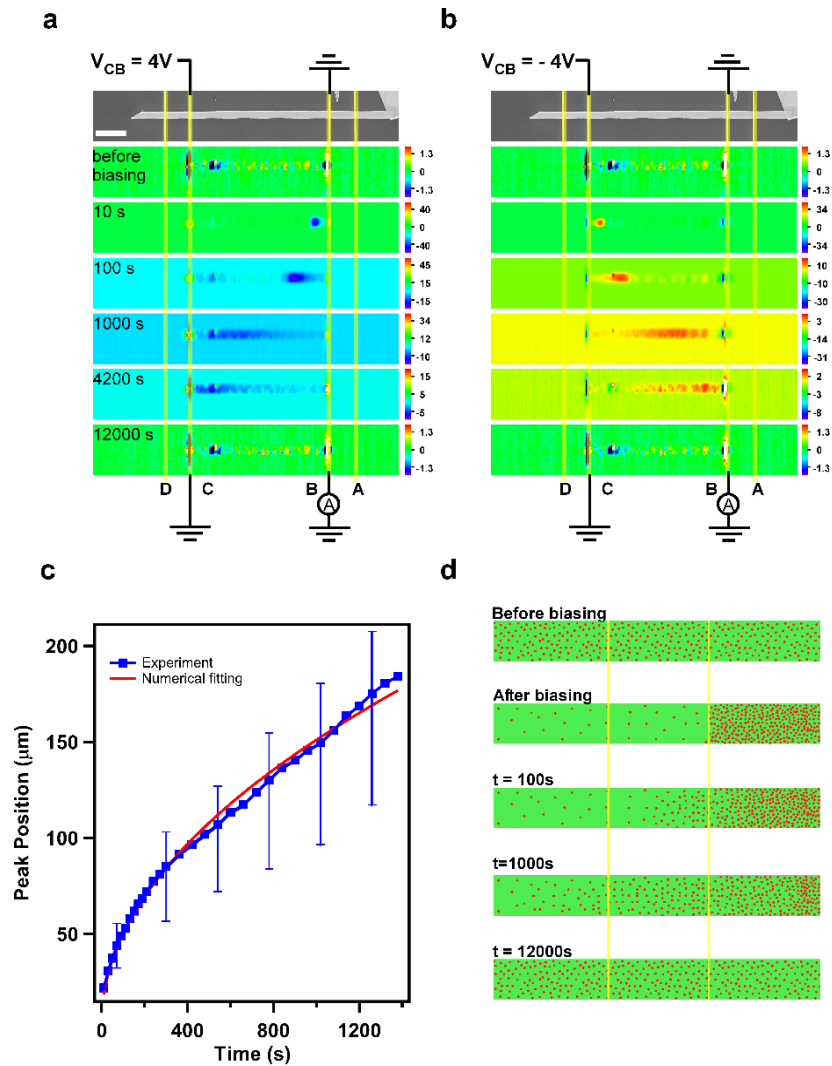

**Fig. 3: Visualization of field-induced doping gradient in a TI nanoribbon by photocurrent mapping.** All photocurrent images were taken at 200 K and zero-bias for device #1. **a,** Optical image and photocurrent images taken at different times. The first photocurrent image was taken after the device was cooled to 200 K in the absence of electric field. Then $V_{CB}$ = 4 V was held for 3 hrs. The rest photocurrent images were taken at the indicated times after $V_{CB}$ was reduced to 0 V at t = 0. The laser wavelength was 532 nm and its power was 50 $\mu$W. Color scales are in nA. Scale bar denotes 50 $\mu$m. **b,** Similar measurements performed for $V_{CB}$ = - 4 V. **c,** Photocurrent peak position in **a** as a function of time. The position of contact B is taken as origin. The red curve is square root fitting. The "error bars" at representative data points indicate the full width at half maximum (FWHM) of the photocurrent peaks. **d**, Schematic drawings of the distribution of trapped holes at each stage.

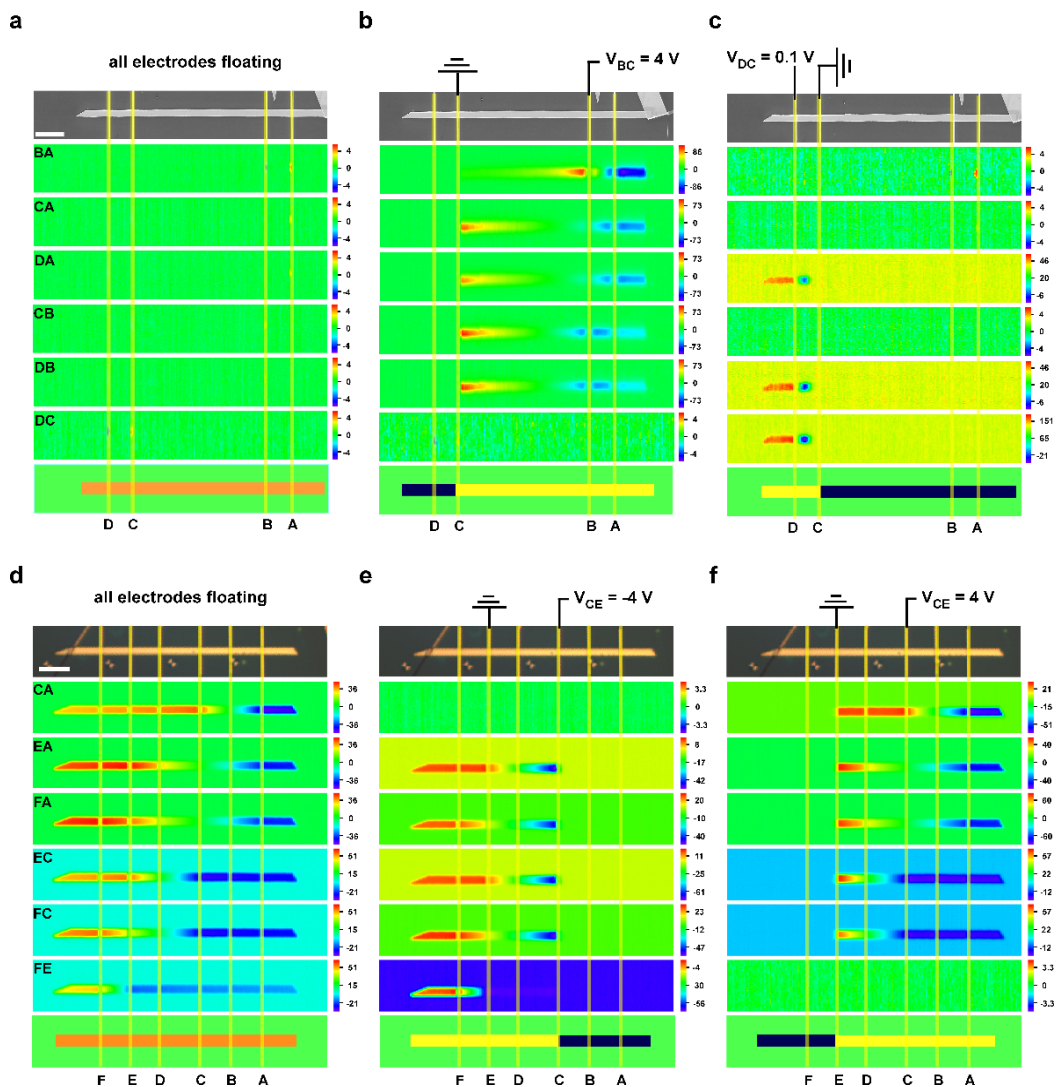

**Fig. 4: Field effects on photocurrent decay length.** Device #1 was cooled from room temperature to 77 K under different biasing conditions as specified on the optical images at the top. All photocurrent images were taken at 77 K and zero-bias. **a-c**, Device #1. Scale bar denotes 50 μm and laser power is 2.3 μW. **d-f**, Device #3. Scale bar denotes 30 μm and laser power is 0.50 μW. Photocurrent images were taken with different connections as indicated in each row. For example, BA means that contact B is directly grounded and A is connected to ground through a preamplifier. The bottom row is a schematic drawing indicating the field-induced conductivity change: orange, no change; yellow, less *n*-type; dark blue, more *n*-type. Color scales are in nA.

# References

1      Kondou, K. *et al.* Fermi-level-dependent charge-to-spin current conversion by Dirac surface states of topological insulators. *Nat. Phys.* **12**, 1027-1031 (2016).

2      Peng, X., Yang, Y., Singh, R. R. P., Savrasov, S. Y. & Yu, D. Spin generation via bulk spin current in three-dimensional topological insulators. *Nat. Commun.* **7**, 10878 (2016).

3      Fu, L. & Kane, C. L. Superconducting Proximity Effect and Majorana Fermions at the Surface of a Topological Insulator. *Phys. Rev. Lett.* **100**, 096407 (2008).

4      Linder, J., Tanaka, Y., Yokoyama, T., Sudbø, A. & Nagaosa, N. Unconventional Superconductivity on a Topological Insulator. *Phys. Rev. Lett.* **104**, 067001 (2010).

5      Qu, D.-X. *et al.* Onset of a Two-Dimensional Superconducting Phase in a Topological-Insulator-- Normal-Metal $Bi_{1-x}Sb_x$/Pt Junction Fabricated by Ion-Beam Techniques. *Phys. Rev. Lett.* **121**, 037001 (2018).

6      Hou, Y. *et al.* Millimetre-long transport of photogenerated carriers in topological insulators. *Nat. Commun.* **10**, 5723 (2019).

7      Pertsova, A. & Balatsky, A. V. Dynamically Induced Excitonic Instability in Pumped Dirac Materials. *Ann. Phys.* **532**, 1900549 (2020).

8      Kong, D. *et al.* Ambipolar field effect in the ternary topological insulator $(Bi_xSb_{1-x})_2Te_3$ by composition tuning. *Nat. Nanotechnol.* **6**, 705-709 (2011).

9      Hsieh, D. *et al.* A tunable topological insulator in the spin helical Dirac transport regime. *Nature* **460**, 1101-1105 (2009).

10     King, P. D. C. *et al.* Large Tunable Rashba Spin Splitting of a Two-Dimensional Electron Gas in $Bi_2Se_3$. *Phys. Rev. Lett.* **107**, 096802 (2011).

11     Rischau, C. W., Leridon, B., Fauqué, B., Metayer, V. & van der Beek, C. J. Doping of $Bi_2Te_3$ using electron irradiation. *Phys. Rev. B* **88**, 205207 (2013).

12     Suh, J. *et al.* Simultaneous Enhancement of Electrical Conductivity and Thermopower of $Bi_2Te_3$ by Multifunctionality of Native Defects. *Adv. Mater.* **27**, 3681-3686 (2015).

13     Zhu, Z. H. *et al.* Rashba Spin-Splitting Control at the Surface of the Topological Insulator $Bi_2Se_3$. *Phys. Rev. Lett.* **107**, 186405 (2011).

14     Okada, Y. *et al.* Ripple-modulated electronic structure of a 3D topological insulator. *Nat. Commun.* **3**, 1158 (2012).

15     Kim, D. *et al.* Surface conduction of topological Dirac electrons in bulk insulating $Bi_2Se_3$. *Nat. Phys.* **8**, 459-463 (2012).

16     Checkelsky, J. G., Hor, Y. S., Cava, R. J. & Ong, N. P. Bulk Band Gap and Surface State Conduction Observed in Voltage-Tuned Crystals of the Topological Insulator $Bi_2Se_3$. *Phys. Rev. Lett.* **106**, 196801 (2011).

17     Chen, J. *et al.* Gate-Voltage Control of Chemical Potential and Weak Antilocalization in $Bi_2Se_3$. *Phys. Rev. Lett.* **105**, 176602 (2010).

18     Kong, D. *et al.* Rapid Surface Oxidation as a Source of Surface Degradation Factor for $Bi_2Se_3$. *ACS Nano* **5**, 4698-4703 (2011).

19     Benia, H. M., Lin, C., Kern, K. & Ast, C. R. Reactive Chemical Doping of the $Bi_2Se_3$ Topological Insulator. *Phys. Rev. Lett.* **107**, 177602 (2011).

20     Yeats, A. L. *et al.* Local optical control of ferromagnetism and chemical potential in a topological insulator. *Proc. Natl. Acad. Sci. U.S.A.* **114**, 10379-10383 (2017).

21      Yeats, A. L. *et al.* Persistent optical gating of a topological insulator. *Sci. Adv.* **1**, e1500640 (2015).

22      Pan, F., Gao, S., Chen, C., Song, C. & Zeng, F. Recent progress in resistive random access memories: Materials, switching mechanisms, and performance. *Mater. Sci. Eng. R Rep.* **83**, 1-59 (2014).

23      Sup Choi, M. *et al.* Controlled charge trapping by molybdenum disulphide and graphene in ultrathin heterostructured memory devices. *Nat. Commun.* **4**, 1624 (2013).

24      Bertolazzi, S., Krasnozhon, D. & Kis, A. Nonvolatile Memory Cells Based on $MoS_2$/Graphene Heterostructures. *ACS Nano* **7**, 3246-3252 (2013).

25      Yang, Y. *et al.* Hot Carrier Trapping Induced Negative Photoconductance in InAs Nanowires toward Novel Nonvolatile Memory. *Nano Lett.* **15**, 5875-5882 (2015).

26      Wang, Q. *et al.* Nonvolatile infrared memory in $MoS_2$/PbS van der Waals heterostructures. *Sci. Adv.* **4**, eaap7916 (2018).

27      Tarun, M. C., Selim, F. A. & McCluskey, M. D. Persistent Photoconductivity in Strontium Titanate. *Phys. Rev. Lett.* **111**, 187403 (2013).

28      Bhatnagar, A., Kim, Y. H., Hesse, D. & Alexe, M. Persistent Photoconductivity in Strained Epitaxial $BiFeO_3$ Thin Films. *Nano Lett.* **14**, 5224-5228 (2014).

29      Yin, H., Akey, A. & Jaramillo, R. Large and persistent photoconductivity due to hole-hole correlation in CdS. *Phys. Rev. Mater.* **2**, 084602 (2018).

30      Yashina, L. V. *et al.* Negligible Surface Reactivity of Topological Insulators $Bi_2Se_3$ and $Bi_2Te_3$ towards Oxygen and Water. *ACS Nano* **7**, 5181-5191 (2013).

31      Hong, S. S., Cha, J. J., Kong, D. & Cui, Y. Ultra-low carrier concentration and surface-dominant transport in antimony-doped $Bi_2Se_3$ topological insulator nanoribbons. *Nat. Commun.* **3**, 757 (2012).

32      Hou, Y., Xiao, R., Tong, X., Dhuey, S. & Yu, D. In Situ Visualization of Fast Surface Ion Diffusion in Vanadium Dioxide Nanowires. *Nano Lett.* **17**, 7702-7709 (2017).

33      Kastl, C. *et al.* Local photocurrent generation in thin films of the topological insulator $Bi_2Se_3$. *Appl. Phys. Lett.* **101**, 251110 (2012).

34      Kastl, C. *et al.* Chemical potential fluctuations in topological insulator $(Bi_{0.5}Sb_{0.5})_2Te_3$ -films visualized by photocurrent spectroscopy. *2D Mater.* **2**, 024012 (2015).

35      Fanetti, M. *et al.* Growth, morphology and stability of Au in contact with the $Bi_2Se_3$(0 0 0 1) surface. *Appl. Surf. Sci.* **471**, 753-758 (2019).

36      Shaughnessy, M. C., Bartelt, N. C., Zimmerman, J. A. & Sugar, J. D. Energetics and diffusion of gold in bismuth telluride-based thermoelectric compounds. *J. Appl. Phys.* **115**, 063705 (2014).

37      Kim, T.-H. *et al.* Phase Transformation of Alternately Layered Bi/Se Structures to Well-Ordered Single Crystalline $Bi_2Se_3$ Structures by a Self-Organized Ordering Process. *J. Phys. Chem. C* **116**, 3737-3746 (2012).

38      Kim, Y. S. *et al.* Thickness-dependent bulk properties and weak antilocalization effect in topological insulator $Bi_2Se_3$. *Phys. Rev. B* **84**, 073109 (2011).

39      Shewmon, P. *Diffusion in Solids.*    (Springer, Cham, Swizerland, 2016).

40      Xu, Y. *et al.* Disorder enabled band structure engineering of a topological insulator surface. *Nat. Commun.* **8**, 14081 (2017).

41      Wang, R., Erten, O., Wang, B. & Xing, D. Y. Prediction of a topological p + ip excitonic insulator with parity anomaly. *Nat. Commun.* **10**, 210 (2019).

**Acknowledgements** This work was supported by National Science Foundation Grant DMR-1838532 and DMR-1710737. This research used the Molecular Foundry, which is a US Department of Energy Office of Science User Facility under contract no. DE-AC02-05CH11231. We acknowledge Luke McClintock, H. Clark Travaglini, Antonio Rossi, Inna Vishik, Giacomo Resta, and Sergey Savrasov for assistance in experiment and discussion.

**Author contribution** D. Y. and Y. H. designed the experiments. Y. H. synthesized $Bi_2Se_3$ nanoribbons and performed the measurements. R. X. and Y. H. performed numerical simulation. S. L. and L. W. assisted the synthesis and measurements. D. Y. and Y. H co-wrote the paper.

## Methods

**Nanoribbon Growth and Device Fabrication.** The CVD growth was carried out in a Lindberg Blue M tube furnace, following a similar procedure as in our previous work[29]. $Bi_2Se_3$ powder (99.999%, Alfa Aesar) mixed with Sb powder (99.999%, Alfa Aesar) was placed at the centre of the tube furnace. Se pellets (99.999%, Johnson Matthey Inc.) were placed upstream at a distance of 16 cm. A silicon substrate coated with a 10 nm Au film was placed downstream 14 cm from the centre of the furnace. Argon gas was used to transfer vapour from the source materials to the growth substrate at a 150 sccm (standard cubic centimetres per minute) flow rate. During the 5 hours of growth time, the temperature at the centre of the furnace was maintained at 680 °C and the pressure was maintained at room pressure. The furnace was then cooled down to room temperature over approximately 3 hours. The as-grown nanoribbons were transferred to Si substrates covered by 300 nm $SiO_2$, where single nanoribbon field effect transistor (FET) devices were fabricated following the standard electron beam lithography process. Top metal contacts (10 nm Ti / 290 nm Au) were made using an electron beam evaporator (CHA). A typical device is shown in Fig. 1b.

**Optoelectronic measurements.** All measurements were carried out in a cryostat (Janis ST-500) under $10^{-7}$ Torr. Current-voltage curves were measured through a current preamplifier (DL Instruments, model 1211) and a NI data acquisition system. Scanning photocurrent microscopy (SPCM) measurements were performed using a home-built setup based upon an Olympus microscope. A 532 nm CW laser was focused by a 10× NA = 0.3 objective lens to an approximately 3-μm spot and raster scanned on a planar nanoribbon device by a pair of mirrors mounted on galvanometers. Both reflectance and photocurrent were simultaneously recorded to produce a 2D maps. The laser power was controlled by a set of neutral density (ND) filters and was measured by a power meter underneath the objective lens. The general trends are highly repeatable in more than 10 devices measured to date.

**Diffusion model.** Simulation is carried out by solving a 1D diffusion-drift equation:

$$\frac{dn}{dt} = \mu E \frac{dn}{dx} - D \frac{d^2n}{dx^2} \tag{1}$$

Here, $n$ is the concentration of localized charges, $E$ is the electric field, $\mu$ is their mobility, and $D$ is their diffusion coefficient (Einstein relation $D = \mu k_B T/q$ is assumed). Because of a small contact resistance (< 8% of the channel resistance), the electric field is small outside the channel but not exactly zero. The partial differential equation is solved by finite element method with a mesh size of 1 μm. The time step used in the simulation is 0.2 ms above 200K and 2.5 ms below. Charge continuity and current continuity are applied at all boundaries. A uniform trapped hole density is assumed as the initial condition $n\ (t = 0) = n_0$. Temporal evolution of $n$ under $V_{AB} = \pm 0.5\ V$ are presented in Extended Data Fig. 7. We notice that as $G_{CD}$ is close to saturation, its evolution is slightly slower than predicted by the diffusion model (Fig. 1d) likely because the repulsive interaction between the trapped charges which is not considered in the model and tends to reduce the diffusion rate[42].

42   Kühne, M. *et al.* Ultrafast lithium diffusion in bilayer graphene. *Nat. Nanotechnol.* **12**, 895-900 (2017).

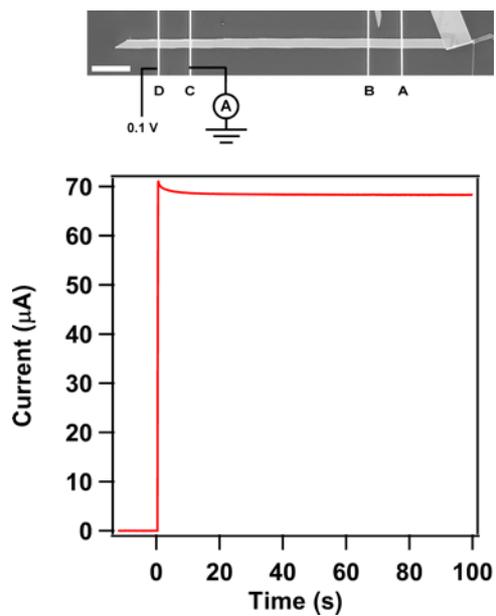

**Extended Data Fig. 1:** Current in channel CD gradually changed as $V_{CD}$ = 0.1 V was maintained at room temperature. Scale bar denotes 50 μm.

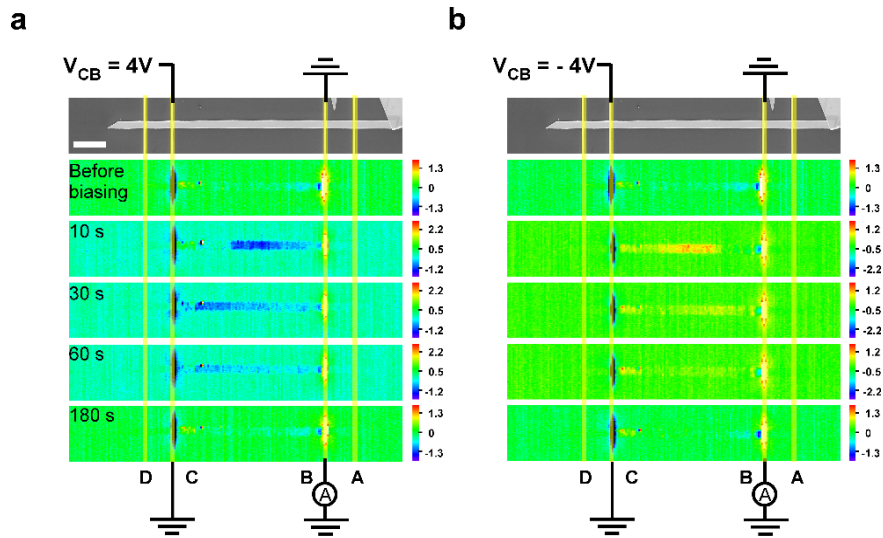

**Extended Data Fig. 2: Visualization of field-induced doping gradient in a TI nanoribbon by photocurrent mapping at 300 K. a**, The first photocurrent image was taken in the absence of electric field. $V_{CB}$ = 4 V was then held for 10 minutes to create doping gradient. The rest photocurrent images were taken at the indicated times after $V_{CB}$ was reduced to 0 V at t = 0. **b**, Similar measurements but with $V_{CB}$ = - 4 V. The laser wavelength was 532 nm and its power was 50 μW. Color scales are in nA. Scale bar denotes 50 μm.

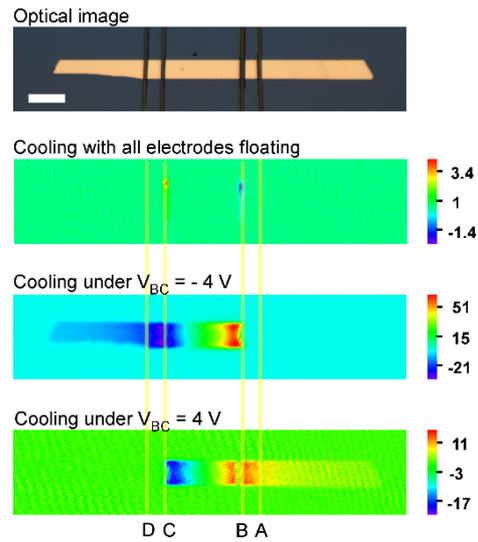

**Extended Data Fig. 3: Field cooling effects on a device with only Ti electrodes (Au-free).**

Photocurrent profiles were measured at 77 K with contact C connected to ground through an ammeter and contact B directly grounded. Long photocurrent decay lengths were achieved, indicating $E_F$ was tuned close to the Dirac point after cooling under field. This experiment excludes the possibility that Au migration causes the observed field cooling effects. Laser power is 10 µW. Colour scale is in nA. Scale bar denotes 50 µm.

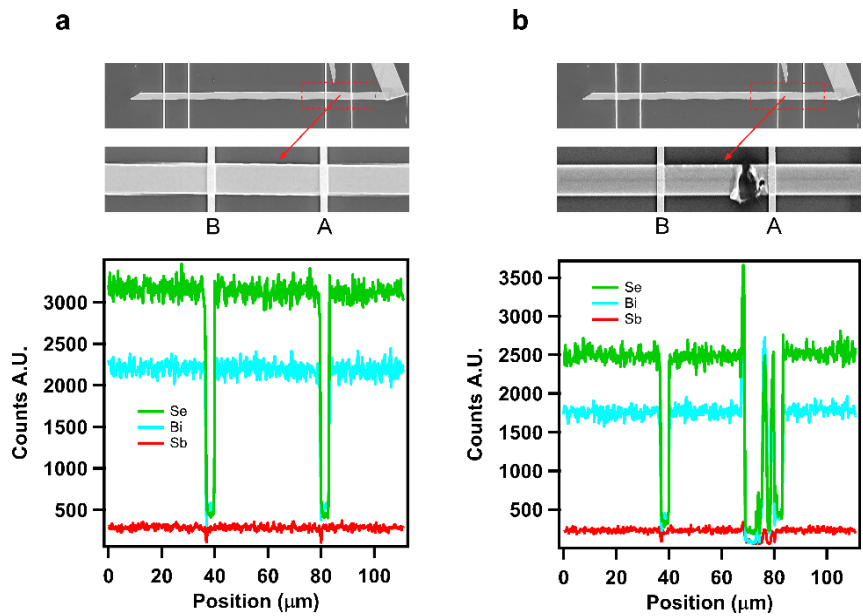

**Extended Data Fig. 4: Control experiment to evaluate the possibility of field-induced ion migration.** A large bias was maintained between contacts A and B ($V_{AB}$ = -2 V) in a $Bi_{2-x}Sb_xSe_3$ nanoribbon device (device #1) at 300 K for 10 min to induce substantial chemical potential shift. Then $V_{AB}$ was further increased to -30 V to intentionally break the channel to prohibit the ion diffusion across the breaking point. If ions (such as excess surface Se or Se vacancies) migrate under field, unequal chemical compositions are expected in the two separate segments. The device was then examined by spatially resolved energy dispersive X-ray spectroscopy (EDS). Chemical composition distributions along the nanoribbon axis did not show observable change ($\sigma_n$ / $n$ < 3%, where $n$ is the number of counts for Se and $\sigma_n$ is its standard deviation), indicating the ions do not migrate within this uncertainty.

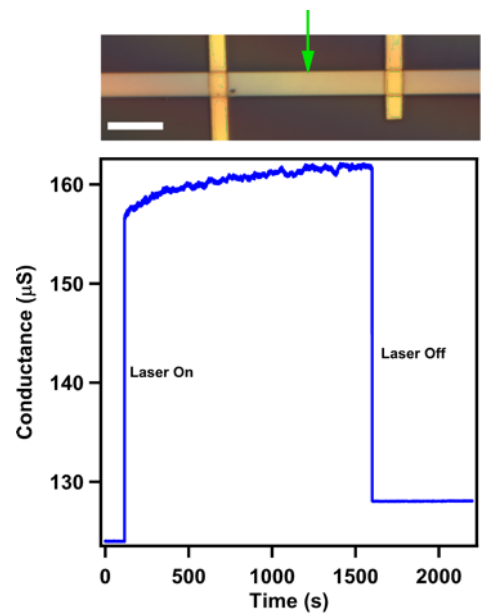

**Extended Data Fig. 5: Photo-induced persistent conductance change at 77 K.** A 500 μW

laser at 532 nm focused by a 10× objective lens led to a sharp conductance jump, then followed by a slow increase. After the laser was turned off, conductance was increased by 3% compared to the initial value. Scale bar denotes 10 μm. Green arrow indicates the laser position.

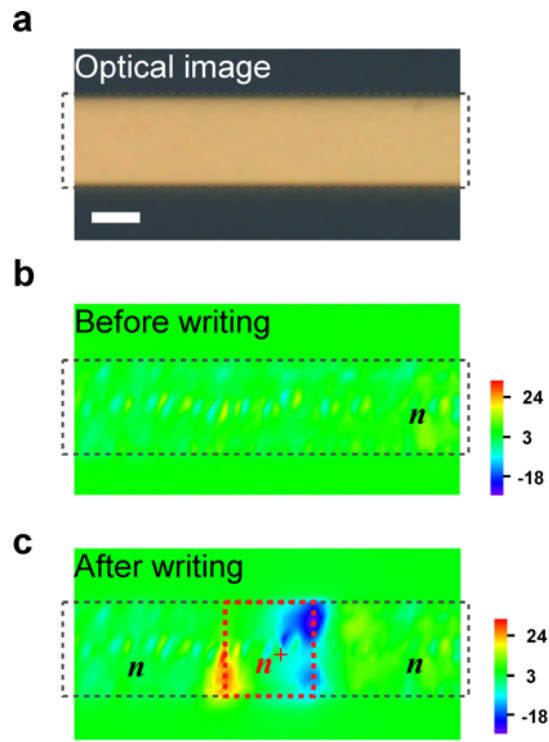

**Extended Data Fig. 6: Optically induced $E_F$ shift. a**, Optical image of a segment of a nanoribbon device. Scale bar denotes 10 μm. **b**, **c**, SPCM images taken at 77 K before and after scanning a high-power laser over the area indicated by the red dashed box in **c**. Colour scale is in nA. The laser used for optical writing was at 1.5 mW and the writing took 30 mins. The laser power for SPCM was 10 μW. Photocurrent spots with opposite polarities were observed at the boundaries of the writing area, indicating formation of $n$-$n^+$ junctions at the edges.

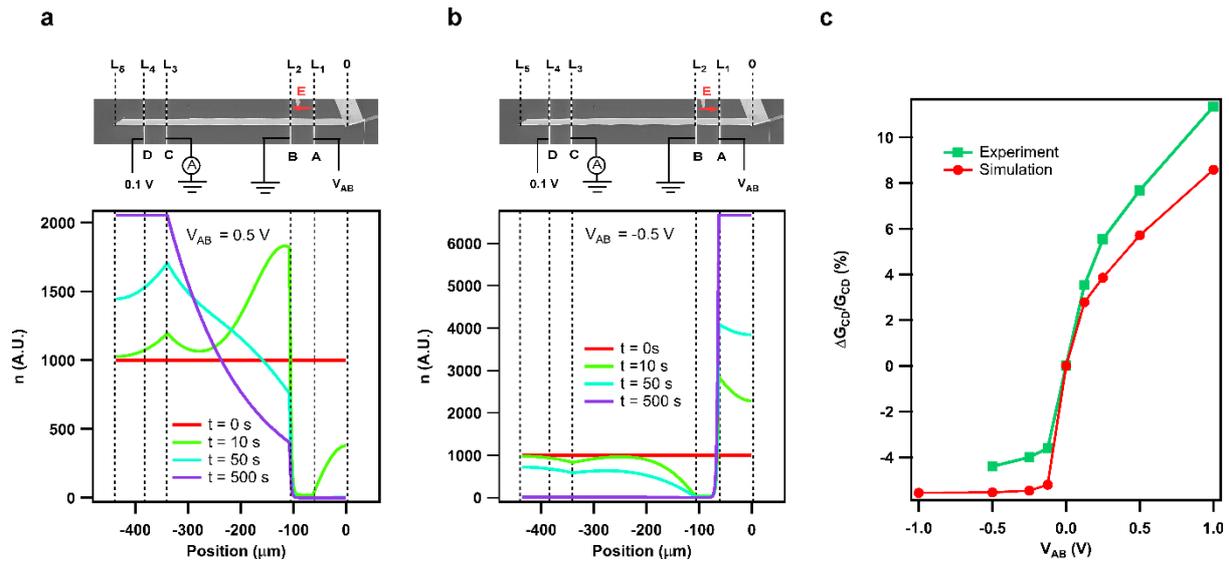

**Extended Data Fig. 7: Numerical modeling of electric-field-driven trapped charge migration. a, b,** Simulated evolution of the trapped hole distribution $n(x,t)$ after turning on $V_{AB} = \pm 0.5\ V$ at $t = 0$, respectively. The circuit connections were shown in the SEM images on the top. **c,** The percentage change of conductance (extracted from Fig. 1c) as a function of $V_{AB}$ at 300K. The simulation follows well the experimental results.

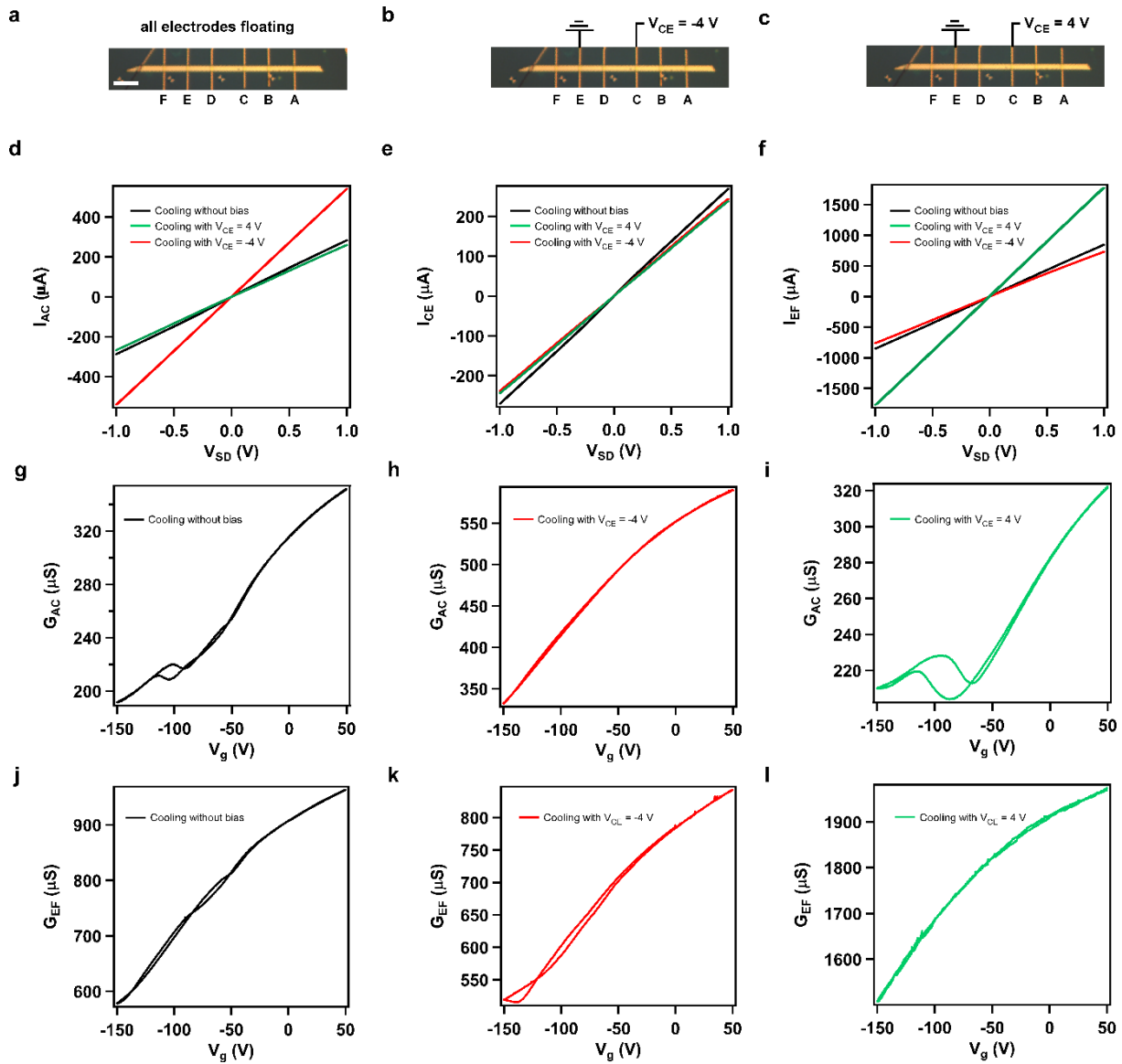

**Extended Data Fig. 8: Field cooling effects on gate responses for device #3. a-c,** Optical image and schematics showing three biasing conditions. Scale bar denotes 30 μm. **d-f,** I-$V_{sd}$ curves in different channels after cooling with different biasing conditions. **g-l,** Gate responses of channels AC (**g-i**) and EF (**j-l**) at 77 K, respectively, after cooling under different biasing conditions. The non-monotonic changes of conductance in **g**, **i**, and **k** indicate $E_F$ is tuned to the Dirac point.

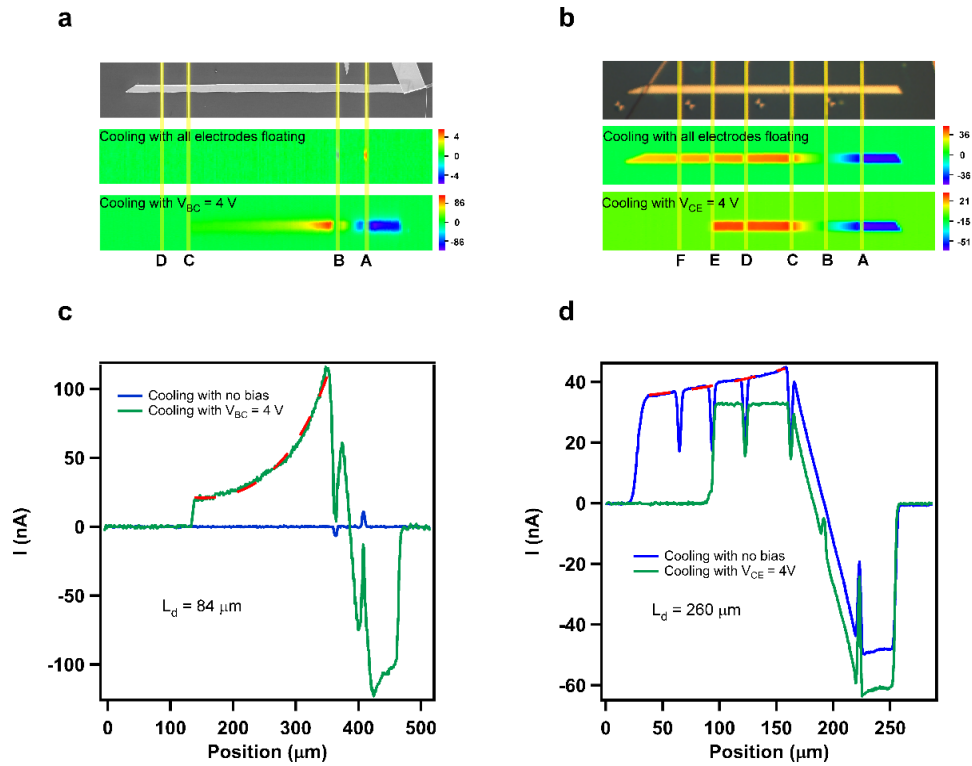

**Extended Data Fig. 9: Field effects on photocurrent cross sections at 77 K. a, b,** Device images and photocurrent images under different biasing conditions as shown in Fig. 4. **c, d,** Photocurrent cross sections extracted from **a** and **b**. Photocurrent decay lengths ($L_d$) were obtained by fitting with a cosh(-x/$L_d$) function. **a, c,** device #1. **b, d,** device #3.